\documentclass[manuscript]{aastex}

\shorttitle{The tail of the turtle}
\shortauthors{Marelli et al.}

\begin{document}

\title{PSR J0357+3205: the tail of the turtle}

\author{M. Marelli,\altaffilmark{1} A. De Luca,\altaffilmark{1,2} D. Salvetti,\altaffilmark{1,2,3} N. Sartore,\altaffilmark{1}
A. Sartori,\altaffilmark{1,2,3} P. Caraveo\altaffilmark{1,2}, F. Pizzolato\altaffilmark{1}, P.M. Saz Parkinson\altaffilmark{4}
and A. Belfiore\altaffilmark{1,4}}

\altaffiltext{1}{INAF - Istituto di Astrofisica Spaziale e Fisica Cosmica Milano, via E. Bassini 15, 20133 Milano, Italy}
\altaffiltext{2}{Istituto Nazionale di Fisica Nucleare, Sezione di Pavia, 
Via Bassi 6, I-27100 Pavia (Italy)}
\altaffiltext{3}{Universit\`a degli Studi di Pavia, Strada Nuova 65, 27100 Pavia, Italy}
\altaffiltext{4}{Santa Cruz Institute for Particle Physics, University of California, Santa Cruz, CA 95064}
\email{marelli@iasf-milano.inaf.it}

\begin{abstract}
Using a new {\it XMM-Newton} observation, we have characterized the
X-ray properties of the middle-aged radio-quiet $\gamma$-ray pulsar
J0357+3205 (named Morla) and its tail. The X-ray emission from the
pulsar is consistent with a magnetospheric non-thermal origin plus a 
thermal emission from a hot spot (or hot spots). The lack of a thermal
component from the whole surface makes Morla the coldest neutron star in its 
age range. We found marginal evidence for a double-peaked modulation 
of the X-ray emission. The study of the 9'-long tail confirmed the lack 
of extended emission near the pulsar itself. The tail shows a very asymmetric 
brightness profile and its spectrum lacks any spatial variation. We
found the nebular emission to be inconsistent with a classical
bow-shock, ram-pressure dominated pulsar wind nebula. We propose
thermal bremsstrahlung as an alternative mechanism for Morla's tail
emission. In this scenario, the tail emission comes from the
shocked interstellar medium (ISM) material heated up to X-ray
temperatures. This can fully explain the peculiar features of the
tail, assuming a hot, moderately dense interstellar medium around the
pulsar. For a bremsstrahlung-emitting tail, we can estimate the pulsar
distance to be between 300 and 900 pc. A pulsar velocity of $\sim$1900
km s$^{-1}$ is required - which would make Morla the pulsar with the largest
velocity - and high inclination angles ($>$70$^{\circ}$) are
preferred. We propose Morla's nebula as the first example of a new
``turtle's tail'' class of thermally-emitting nebulae associated to high
velocity pulsars.
\end{abstract}

\keywords{Stars: neutron --- Pulsars: general --- Pulsars: individual (PSR
  J0357+3205) --- X-rays: stars}


\section{Introduction} \label{intro}

The dramatic increase in the number of known gamma-ray pulsars since the launch of the {\it Fermi Gamma-ray Space
Telescope} offers the first opportunity to study a sizeable population of these high-energy sources.
The {\it Fermi} Large Area Telescope (LAT, \citet{atw09}) has discovered pulsed $\gamma$-ray signals
from more than 100 objects \citep{abd10a,nol13}, revolutionizing our view of them and
marking the birth of new high-energy pulsar sub-families, such as millisecond \citep{abd09a} and radio-quiet
$\gamma$-ray pulsars \citep{abd09b}, that are as numerous as the
classical one. The wealth of detections confirms the importance of the $\gamma$-ray channel
in the overall energy budget of rotation-powered pulsars and
points to emission models in which the $\gamma$-ray
production occurs in the outer magnetosphere along open-field lines
(outer gap/slot gap, \citet{che86,har04}) and paves the way for our
understanding of the three-dimensional structure and electrodynamics of neutron star magnetospheres.

In a sense, the most important objects to constrain pulsar models are
the ``extreme'' ones, accounting for the tails of the population distribution in energetics, age and magnetic field.
In this respect, PSR J0357+3205 is one of the most interesting pulsars
discovered by the LAT. J0357+3205, which we name ``Morla"
(cf. ``The Neverending Story", M. Ende) for its slow rotation among
{\it Fermi}-LAT pulsars (P=444 ms), was discovered in a blind search during the
first three months of {\it Fermi}-LAT scanning \citep{abd09c}. It is located off the Galactic plane, at a latitude of b$\simeq$-16$^{\circ}$.
While Morla is not the oldest rotation-powered pulsar ($\tau_c$=5.4
$\times$ 10$^5$ yr) - with an age similar to the well-known ``Three
Musketeers'' \citep{del05} - it is the slowest rotator among {\it
Fermi} pulsars. Its spin-down luminosity, as low as
$\dot{E}_{rot}$ = 5.9 $\times$ 10$^{33}$ erg s$^{-1}$, makes it one of the non-recycled $\gamma$-ray pulsars
with the smallest rotational energy loss detected so far. This suggests that PSR J0357+3205 is quite nearby:
by scaling its $\gamma$-ray flux using the ``pseudo-distance'' relation from \citet{saz10}, we obtain a distance of $\sim$500 pc,
making Morla a natural target for X-ray observations.

Exploiting a joint {\it Chandra} and {\it NOAO} program, \citet{del11}
found the X-ray counterpart of the $\gamma$-ray pulsar to be located 
at R.A.(J2000) = 03$^h$57$^m$52$^s$.32, Dec.(J2000) = +32$^{\circ}$05'20.6"
(0.25" error radius, 68\% confidence level). The pulsar X-ray emission was found to be consistent with a purely
magnetospheric, non-thermal origin leading to a value of the $\gamma$-to-X flux ratio of $\sim$1000, well below the mean
for radio-quiet pulsars ($\sim$9000, \citet{mar12}). Unfortunately,
the poor statistics precluded a clear detection of a thermal component
in the neutron star spectrum as well as a full characterization of its column density and distance.
Nevertheless, the {\it Chandra} data unveiled the presence of an extended structure
protruding from the pulsar and extending $>$9' in
length, which would correspond to $\sim$1.3 pc for a pulsar distance of
500 pc, and $\sim$1.5' in width. The ``tail" is apparently detached from
the pulsar, without detection of nebular counts until 50"
from the pulsar; it shows an unusual, asymmetric morphology, with the
brightnest part far from the pulsar. {\it NOAO} optical images, taken
in V and K bands, show no hints of correlated, diffuse emission. No
extended source was detected either in public radio or infrared surveys.
The lack of any H$\alpha$ emission around the pulsar \citep{del12} points
to a full ionized, likely hot ($>10^4$ K), ISM.\\
Recently, using a new {\it Chandra} observation, \citet{del12}
detected the pulsar's proper motionof 165$\pm$30 mas yr$^{-1}$, in the direction away from the tail
and parallel to the Galactic plane. Assuming a
pulsar distance of 500 pc, this results in a lower limit velocity of
390 km s$^{-1}$, making it a relatively high-velocity pulsar \citep{man05}.
Many extended tails of X-ray emission have been discovered, 
associated with a number of rotation powered pulsars
(e.g. \citet{kar08}) within the framework of a ram-pressure dominated pulsar wind nebula (see \citet{gae06}
for a review). However, the unusual phenomenology of the feature we
observed is barely consistent with such a picture. There is no hint of
diffuse emission in the pulsar surroundings, where bright emission from
the wind termination shock should be seen (as observed in other known cases).
Moreover, the asymmetric brightness profile of the tail, with its 
sharp north-eastern edge and its broad maximum at large distance 
from the pulsar is remarkably different from any other observed
structure (with the possible exception of the peculiar feature
associated with PSR B2224+65, powering the Guitar nebula - see \citet{hui07}).
The mere existence of the tail is also problematic for a pulsar with
such a low $\dot{E}_{rot}$ as J0357+3205 (see e.g. \citet{jag08}).

In order to assess the nature of the tail and its relationship with the pulsar, 
as well as to better constrain the pulsar emission properties and physics,
we obtained a deep observation of the pulsar and its peculiar tail with {\it XMM-Newton}.
Exploiting its high spectral resolution, we were able to disentangle the
components of the pulsar spectrum and constrain its distance based on the absorption column density.
A complete study of the timing behavior of Morla in X-rays was carried out as well.
Finally, we studied the properties of the tail, in order to characterize
the mechanisms responsible for its emission.

\section{Observations and Data Reduction} \label{obs}

Our deep {\it XMM-Newton} observation of Morla started on 2011 September 15 at 02:37:18 UT
and lasted 111.3 ks. The PN camera \citep{str01} of the EPIC instrument was operating
in Large Window mode (time resolution of 47.7 ms over a 14' $\times$ 27' field of view (FOV)),
while the MOS detectors \citep{tur01} were set in full frame mode (2.6 s time resolution on
a 15' radius FOV). The thin optical filter was used for the PN camera while we chose to use
the medium filter for the MOS detectors due to the presence of moderately bright (m$_R\sim$9) stars.
We used the {\it XMM-Newton} Science Analysis Software (SAS) v11.0. After standard
data processing, using the {\tt epproc} and {\tt emproc} tools, and screening for high particle background time
intervals (following \citet{del04}), the good, dead-time corrected exposure time was 98.5 ks
for the PN and 108.3 ks for the two MOS detectors. In order to fully characterize
both the pulsar and the nebula, we also used the three available {\it Chandra}/ACIS
\citep{gar03} observations of the field, performed on 2009 October 25, 26, and 2011 December 25,
lasting 29.5, 47.1, and 29.0 ks respectively (these data sets were included in the investigations by \citet{del11} and
\citet{del12}). We retrieved ``level 2" data from the {\it Chandra} Science Archive and used the {\it Chandra}
Interactive Analysis of Observation (CIAO) software v.4.2. We also combined the {\it Chandra} spectra,
response matrices and effective area files using the {\tt mathpha}, {\tt addarf} and {\tt addrmf} tools.

\subsection{Imaging and Source Detection} \label{ima}

\begin{figure}
\centering
\includegraphics[angle=0,scale=.80]{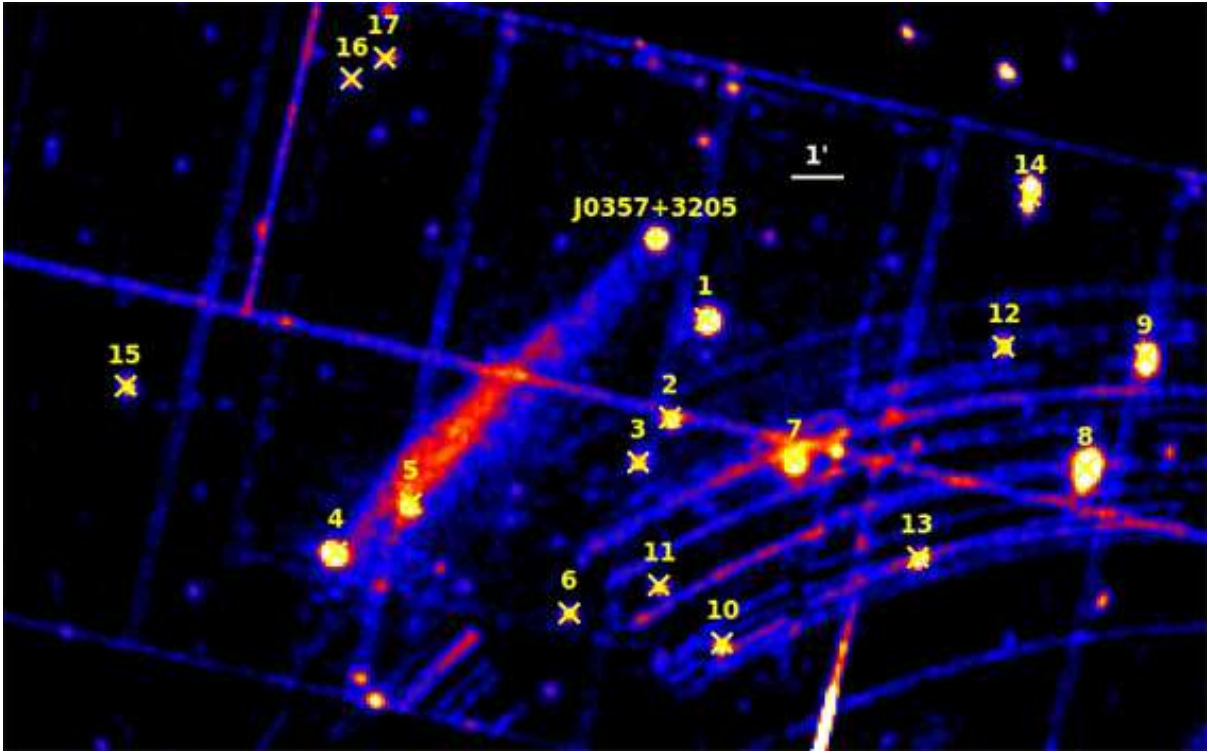}
\caption{Combined 0.3-6 keV exposure-corrected {\it XMM-Newton}
field of view, where the PN and two MOS images were added with {\tt ximage} tool. We applied
a gaussian filter with a kernel radius of 5". The yellow crosses mark the serendipitous sources we analyzed
to constrain the value of the Galactic absorption column (see Table ~\ref{tab-1}). In the lower-right part of the
figure the straylight effect is apparent.\label{fig1}}
\end{figure}

Figure~\ref{fig1} shows the 0.3-6 keV exposure-corrected {\it XMM-Newton}
FOV image, where the PN and two MOS images were added with the {\tt ximage} tool. We applied
a Gaussian filter with a kernel radius of 5". 
The image clearly shows the straylight effect, probably due to the bright source X Persei
that dominates the X-ray sky $\sim$70' to the southwest of Morla, with a flux of $\sim$10$^{-9}$ erg cm$^{-2}$ s$^{-1}$.
X-ray straylight in EPIC is produced by photons which experience single reflection by the mirror
hyperbolae and which reach the sensitive area of the camera and
produce near-annular structures in the image. In order to study the background sources in the
{\it XMM-Newton} observation, we evaluated the spectrum of our straylight annuli.
This is consistent, at low energies, with the X Persei spectrum
found by \citet{lap07}, thus we added and froze these components
in the spectral model of sources affected by this phenomenon.

Source detection using maximum likelihood fitting was done
simultaneously on each of the EPIC-PN, MOS1, and MOS2 using the SAS
tool {\tt edetect\_chain}. This tool runs on the event lists and invokes several other SAS tools to
produce background, sensitivity, and vignetting-corrected exposure
maps. A likelihood threshold of 10 was used for source detection, corresponding to a significance level of 3.6$\sigma$.
Figure~\ref{fig1} shows the brightest sources in the PN field of view; they were analyzed
in order to produce X-ray spectra and find optical counterparts.
The first 7 objects were also observed by Chandra so that we used both the telescope datasets
to obtain radial profiles and spectra.\\
First, we analyzed their radial brightness profile and compared them with
the {\it XMM-Newton} and {\it Chandra} theoretical Point Spread Functions (PSFs).
We found sources 8 and 14 to be the superimposition of two different point-like sources. Using the 2011 Chandra data, source 7
is incompatible with a simple point-like source. This is in agreement
with results from XMM-Newton data - though heavily affected by straylight photons -
of a point-like source surrounded by extended emission. The NASA/IPAC Extragalactic Database (NED \footnote{http://ned.ipac.caltech.edu/})
reports one radio source compatible with the point-like emission, identifying it as an Active Galactic Nucleus (AGN) in a cluster.\\
Each spectrum was then fitted either with an absorbed powerlaw, well-suited for
AGNs as well as pulsars and an absorbed {\tt apec},
well-suited for stellar coronae. Only source 9 requires a double {\tt apec}, quite common
for stellar coronae; for this source we thus cannot exclude the superimposition of two
different stars closer than $\sim$2". Three sources (11, 13, and 14)
cannot be described by the simple models considered here; deeper
studies on the serendipitous sources are beyond the aim of this paper 
so they were excluded from our analysis.\\
Optical counterparts were searched in the deep (4hr 26min exposure) optical and near-infrared
images collected with the 4m Mayall Telescope on 2009 November 10,
using the large-field (36'$\times$36') Mosaic CCD Imager \citep{jac98}.
All the results of our X-ray and optical analyses on background sources are reported in Table~\ref{tab-1}.\\
We found 2 sources (8 and 9) that are compatible with a star interpretation both in the X-ray and
optical bands. These two objects are likely located closer to us than Morla, given their lower value
of the absorption column. Deeper optical observations with different
filters would be required to find their distances from the optical absorption.
All the remaining objects are extragalactic, while source 17 presents
a very high value of the X-ray to optical flux ratio. This can be described as an obscured AGN \citep{mai02}.
It is possible that the emission actually comes from two
different sources (thus lowering the X-ray to optical flux ratio), though
the low statistics prevent us from studying the brightness profile of
the source.\\
All the AGNs were fitted together, linking their
N$_H$ to assess the average Galactic absorption column.
The resulting Galactic value N$_H^{gal}$ = 2.10$\pm$0.09 $\times$ 10$^{21}$ cm$^{-2}$ is
in good agreement with the theoretical one predicted by \citet{dic90}
and \citet{kal05} for objects at the limit of the Galaxy ($\sim$500 pc).

\begin{table}
\begin{center}
\caption{X-ray and optical analysis of background sources\label{tab-1}}
\begin{tabular}{crrrrr}
\tableline\tableline
Object & Coordinates$^a$ & spec$_X$ & N$_H$ (10$^{21}$ cm$^{-2}$) & F$_X$/F$_O$ & type\\
\tableline
1 & 03:57:47.50,+32:03:45.10 & pow & 1.83$\pm$0.20 & 5.0 & AGN\\
2 & 03:57:50.95,+32:01:47.70 & pow/apec & 2.11$_{-1.02}^{+1.20}$/1.36$_{-0.71}^{+0.87}$ & 6.5 & AGN\\
3 & 03:57:53.90,+32:00:58.50 & pow/apec & 1.02$_{-0.97}^{+1.11}$/0.66$_{-0.66}^{+0.46}$ & 9.1 & AGN\\
4 & 03:58:21.95,+31:59:11.65 & pow & 1.87$\pm$0.22 & 1.8 & AGN\\
5 & 03:58:15.25,+32:00:17.35 & pow & 1.91$_{-0.58}^{+0.66}$ & 1.0 & AGN\\
6 & 03:58:00.29,+31:58:01.90 & pow/apec & 2.97$_{-1.03}^{+1.29}$/2.00$_{-0.66}^{+0.91}$ & 1.5 & AGN\\
7 & 03:57:39.35,+32:01:03.00 & pow/apec & 2.68$\pm$0.60/1.20$_{-0.34}^{+0.38}$ & 9.2 & AGN\\
8a & 03:57:12.7,5+32:00:50.10 & apec & 1.25$_{-0.68}^{+0.76}$ & 0.0056 & Star\\
8b & 03:57:12.95,+32:00:42.65 & apec & 0.49$_{-0.22}^{+0.25}$ & 0.0028 & Star\\
9 & 03:57:07.35,+32:02:59.65 & 2apec & 0.21$_{-0.11}^{+0.12}$ & 0.0060 & Star\\
10 & 03:57:46.30,+31:57:24.50 & pow & 1.98$_{-0.92}^{+0.64}$ & 3.6 & AGN\\
12 & 03:57:20.20,+32:03:14.95 & pow/apec & 1.69$_{-0.60}^{+0.54}$/0.98$_{-0.35}^{+0.41}$ & 0.12 & AGN\\
15 & 03:58:40.80,+32:02:27.35 & pow & 1.87$_{-0.50}^{+0.46}$ & 2.8 & AGN\\
16 & 03:58:20.35,+32:08:24.85 & pow & 1.83$_{-1.11}^{+1.27}$ & 1.0 & AGN\\
17 & 03:58:17.45,+32:08:50.20 & pow/apec & 1.73$\pm$0.72/0.94$_{-0.41}^{+0.49}$ & 400 & AGN\\
\tableline
\end{tabular}
\tablenotetext{a}{We report the R.A., Dec. (J2000) coordinates of the optical
counterpart, with a typical error of 0.3". If there was no optical 
counterpart, X-ray coordinates are shown, with a typical systematic error of
2" and 1$\sigma$ statistical error of $\sim$1".}
\tablecomments{X-ray and optical parameters of the brightest sources in the PN field of view; the colors are in a logarithmic scale.
All the errors are at a 90\% confidence level. We use both the optical to X-ray flux ratio
and the X-ray spectral type to separate AGNs from stellar coronal emissions. PSF studies show that sources 8 and 14
result from the superimposition of two sources while source 9 can be a single source. The spectra of sources 11, 13, and 14
cannot be described by a powerlaw nor
{\tt apec} components. Source 17 could be associated with an obscured
AGN \citep{mai02}.}
\end{center}
\end{table}

\section{The Pulsar} \label{pan}

\subsection{Pulsar Spectral Analysis} \label{psp}

\begin{figure}
\centering
\includegraphics[angle=0,scale=.80]{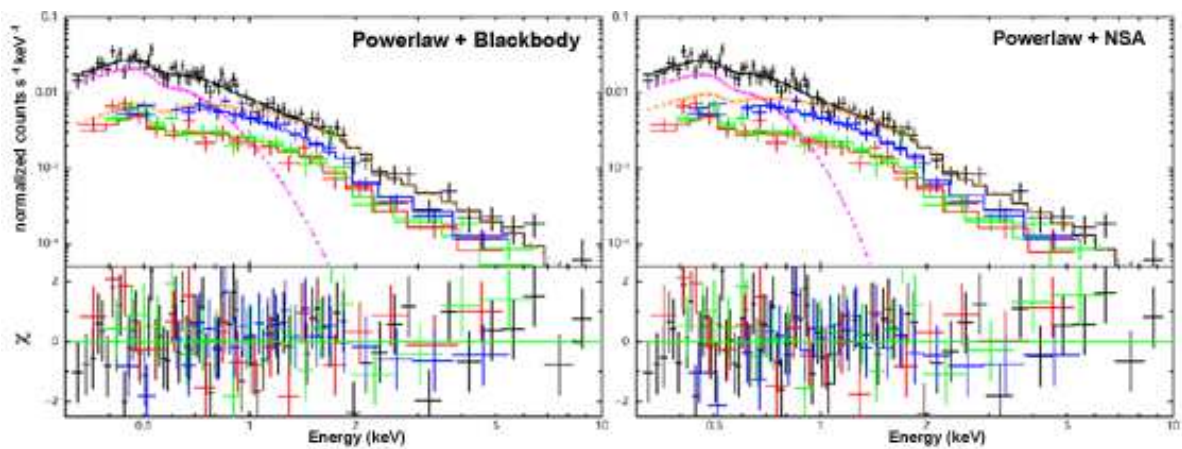}
\caption{PN and MOS spectra of the Morla pulsar. Pattern 0-4 PN events and pattern 0-12 MOS events have been
selected among photons within 18" of the target position. The spectra are rebinned in order to have at least
25 counts per bin. The black, red, and green curves show the data from
the PN and two MOS detectors, respectively. Also shown are the
spectral fits: magenta curves show the thermal components, while orange curves the non-thermal ones.
The data in the left panel are fitted with an absorbed powerlaw plus a
blackbody while the data in the right panel use an absorbed powerlaw plus neutron star atmosphere (nsa) model. The lower
panels show the residuals, in units of standard deviations, with 1 standard deviation error bars.\label{fig2}}
\end{figure}

In order to optimize the pulsar extraction radius we
evaluated the brightness profile of the source to maximize
both the signal-to-noise ratio (S/N) and the number of counts.
The PSF of the EPIC-PN camera onboard {\it XMM-Newton}
is best described by an off-axis, energy-dependent King function \citep{rea04}.
The full width half maximum of the PSF for an on-axis source at 1.5 keV is typically less than 12.5"
for the PN camera and 4.4" for the two MOS detectorss \footnote{http://xmm.esac.esa.int/external/xmm\_user\_support/documentation/sas\_usg/USG/}.
Our results are in agreement with those
published by \citet{del11}, based on {\it Chandra} data,
with no excesses within $\sim30"$ of the pulsar, with the exception
of its bright tail. We found the best extraction radius to be 18",
containing 2022, 515 and 540 counts in the 0.3-10 keV energy range in
the PN and the two MOS detectors respectively,
with a background contribution of $\sim$10\%. We generated an ad hoc response matrix
and effective area files using the SAS tools {\tt rmfgen} and {\tt arfgen}. We also analyzed the
{\it Chandra} spectra of the source using a 1.5" extraction radius (698 pulsar counts
with a background contribution of 0.4\%); we added the three different
spectra with the tool {\tt mathpha}. Response matrix and effective area files were generated
using the CIAO script {\tt psextract}. Due to the presence of the pulsar tail, backgrounds
were evaluated from a circle on the same CCD as the pulsar and without
serendipitous X-ray sources, with a radius of 10" for {\it Chandra}
and 50" for the XMM telescopes.\\
For the pulsar emission, we tried a simple powerlaw (pow), the combination of a powerlaw and
a blackbody (bb) as well as the combination of a powerlaw and a magnetized neutron star
atmosphere model ({\tt nsa} in XSpec - assuming a neutron star with a radius of 13 km, mass of 1.4 M$_{\odot}$ and
a surface magnetic field of 10$^{13}$ G).
While the simple powerlaw gives a fit only marginally acceptable (reduced chi square $\chi^2_{\nu}$=1.22, 134 degrees of freedom (dof)),
both the composite models yield better fits ({\tt nsa}+pow: $\chi^2_{\nu}$=0.96, 131 dof; bb+pow
$\chi^2_{\nu}$=0.97, 132 dof). An F-test performed comparing a simple powerlaw with
the composite spectra results in a chance probability of less than 10$^{-7}$, pointing
to a significant improvement from adding a thermal component.
The pulsar emission is well described by a powerlaw component with a photon
index of $\Gamma$ = 2.28$_{-0.16}^{+0.17}$ (90\% confidence errors) and a blackbody
component with a temperature of kT = 94.0$_{-9.2}^{+12.1}$ eV and an
emitting radius at R = 0.45$_{-0.19}^{+0.31}d_{500}$ km (where $d_{500}$ is the distance of the pulsar in units of $500$ pc), absorbed by a column
density of N$_H$ = 1.20$_{-0.45}^{+0.52}$ $\times$ 10$^{21}$ cm$^{-2}$.
The 0.3-10 keV non-thermal pulsar luminosity is L$_{nth}$ = 1.63$_{-0.28}^{+0.06}d_{500}^2$
$\times$ 10$^{30}$ erg s$^{-1}$, while the thermal one L$_{nth}$ = 1.14$_{-0.20}^{+0.04}d_{500}^2$
$\times$ 10$^{30}$ erg s$^{-1}$. As expected, the {\tt nsa} model yields a lower temperature of 41.1$_{-6.2}^{+8.9}$ eV
and a larger emitting radius of R = 3.77$_{-2.08}^{+4.18}d_{500}$ km,
and a powerlaw component with $\Gamma$ = 2.21$\pm$0.19, absorbed by a column
density of N$_H$ = 1.49$_{-0.53}^{+0.59}$ $\times$ 10$^{21}$ cm$^{-2}$.
Both results are in agreement with the upper limits on the thermal components found by \citet{del11}.
Figure~\ref{fig2} shows the results of the fit for the two different spectral models.

\subsection{Pulsar Timing Analysis} \label{plc}

In order to carry out a proper timing analysis with EPIC data, a correct photon
arrival time must be assigned to each event detected on board.
The transformation from readout sequences
by the EPIC camera to photon arrival times of each photon
is performed by the EPEA (European Photon Event Analyser,
\citet{kus99}). A problem in the EPEA
can produce time jumps in some observations, which have to be
corrected.
In order to identify any possible remaining time jumps, one must neglect both the
barycentering and the randomization of photon arrival times. The time $\Delta$t
between successive events is then calculated and divided by the Frame Time
(FT) of the relevant mode (47.66384 ms for PN Large Window mode).
A time jump is shown to exist when $\Delta$t/FT is different from
an integer by a quantity larger than a tolerance parameter. Only
those time jumps which happen to be an integer multiple of the
relevant FT would be missed by this method (see e.g. \citet{mar12}).\\
Our {\it XMM-Newton} data show that a time jump happened in all the three instruments
between 2011-09-15 16:04:45.822 and 16:10:14.310 UTC,
at about the half of the observation. Such event makes it more
difficult to find the pulsation of Morla, creating a possible
misalignment in the pulsar phase assigned to the photons before and after the jump.
For this reason, we performed careful timing studies in the two
subsets, as well as in the total data set.

\begin{figure}
\centering
\includegraphics[angle=0,scale=.80]{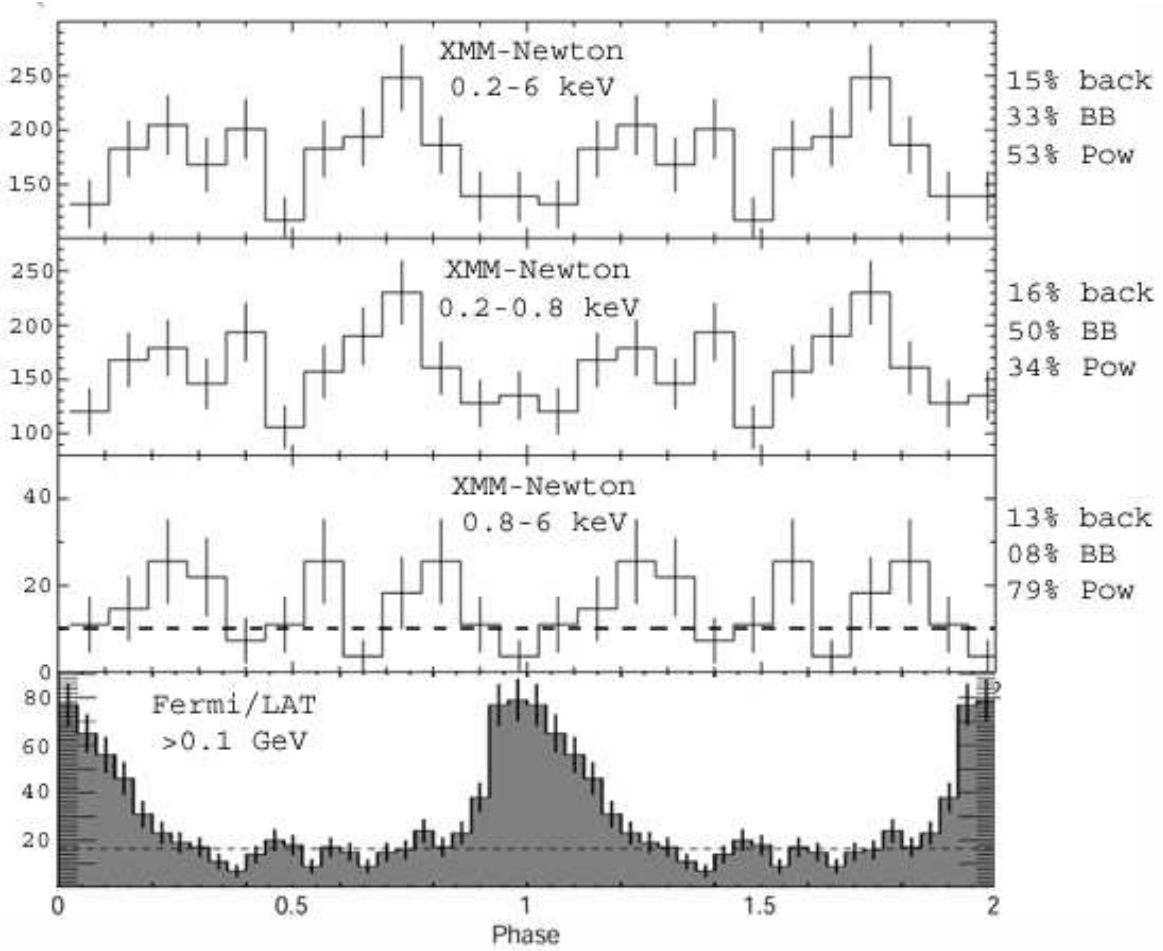}
\caption{EPIC/PN folded light curves in different energy ranges using photons within a 18" radius of the
{\it Chandra} position. X-ray photon phases were computed by folding
at the expected period, based on the most recent available LAT ephemeris: the pulsar period at the start of the observation is
P = 0.44410548661(9)s and the $\dot{P}$ contribution is taken into account. We note that the {\it XMM-Newton} time slip (discussed
in section ~\ref{plc}) could affect our results. Pattern 0 events have been
selected in the 0.2-0.35 keV energy range while Pattern $\leq$ 4 have been used in the 0.35-6 keV range.
For each energy band, the relative contributions of all the spectral components are showen. 
On average the background accounts for $\sim$15\% of the counts. 80\% of the 0.8-6 keV counts come
from the non-thermal pulsar component while 50\% of the 0.2-0.8 keV counts come from the thermal one.
The lower panel shows the LAT light curve of the Morla pulsar from
\citet{abd10a} to which the {\it XMM-Newton} light curves have been aligned in phase.\label{figp}}
\end{figure}

We evaluated the optimal selection of pulsar events by maximizing the signal-to-noise ratio
in the 0.2-6 keV energy range as a function of the source extraction region.
The optimal choice turned out to be a source extraction radius of 18" for the PN camera.
We selected only pattern 0 photons in the range 0.2-0.35 keV
in order to reduce the background contribution due to low-energy electronic noise,
 while in the 0.35-6 keV energy range, we used pattern 0-4 photons.
Photon arrival times were converted to the solar system barycenter
using the SAS tool {\tt barycen}.
According to the accurate {\it Fermi}-LAT ephemeris \citep{nol13},
a period of P = 0.44410548661(9) s is expected at the
epoch of the XMM observation (the uncertainty
on the last digit is computed by
propagating the $3\sigma$ error of the {\it Fermi}-LAT timing solution).
The uncertainty of $\sim10^{-10}$ s is much smaller than the intrinsic
resolution of the {\it XMM-Newton} dataset ($P^2/2\Delta t\sim8.4\times10^{-7}$ s).
Thus, if the spin-down of the pulsar is stable, a single trial folding is appropriate.
Folding using the full LAT ephemeris yields marginal evidence
for a modulation of the X-ray emission (see Figure ~\ref{figp}). A Pearson's $\chi^2$ test
on the folded light curve (10 phase bins) in the 0.2-6 keV 
energy range results in a probability of $5\times10^{-4}$ of sampling a 
uniform distribution. Due to the low signal modulation, a study of
the phase-misalignment before and after the time jump was not possible.
To evaluate the significance of the detection independently of the
phase binning chosen, we computed the corresponding $Z^2$ value 
using the Rayleigh test \citep{buc83} for 1 and 2 harmonics.
This exercise yields probabilities as small as $5\times10^{-4}$ and
$3\times10^{-6}$, respectively, for a chance fluctuation. 
Such a test confirms the hint of a modulation at the pulsar frequency
(and its first harmonic) in the {\it XMM-Newton} data,
with a misalignment between the X-ray and gamma-ray peaks (see Figure~\ref{figp}).
However, the low number of counts prevented us from disentangling the thermal
from the non-thermal pulsation or from performing a phase-resolved spectroscopy analysis.

\section{The Tail} \label{nan}

\subsection{Spatial Analysis} \label{nspa}

The 2009 {\it Chandra} observations revealed an extended, bright nebula protruding from the pulsar, extending
$\sim$9' in length. Unfortunately, the 2011 {\it Chandra} observation does not cover the entire nebular
emission so it cannot be used. The extended feature can be easily studied in our {\it XMM-Newton} observation
due to its lower particle background and higher spectral resolution,
compared to the {\it Chandra} one. On the other hand, the worse
spatial resolution of {\it XMM-Newton} prevented us from significantly constraining the properties of the nebula
near the pulsar. We therefore concentrate on a higher-scale analysis of the extended feature.

\begin{figure}
\centering
\includegraphics[angle=0,scale=.80]{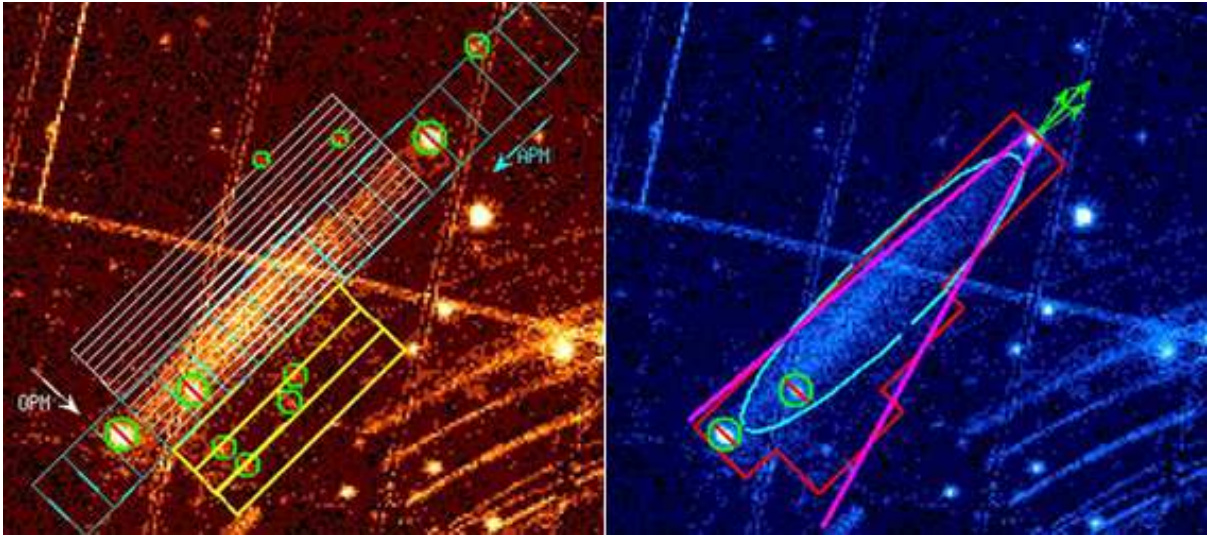}
\caption{0.3-6 keV exposure-corrected {\it XMM-Newton}
view of the tail, where the PN and two MOS images were added with the {\tt ximage} tool.
{\tt Left panel:} The green boxes mark regions used for the brightness profile study along the pulsar proper motion (indicated with an arrow, labeled APM)
while white ones mark regions
used for the brightness profile study orthogonal to the pulsar proper motion (indicated with an arrow, labeled OPM).
Yellow boxes mark the ad-hoc regions used in the region contaminated by
straylight annuli. The barred green circles mark regions
excluded from the analyses due to the presence of point-like sources.
{\tt Right panel:} The red polygon mark the previously defined regions in which the excess due to the presence of the nebula is significant
at 5$\sigma$ or more. Following the red polygon, we traced in magenta a
plausible shape of the tail while the cyan ellipse
is the extraction region used for our spectral analyses on the tail. The long green arrow indicates the direction of the
pulsar proper motion, while the two shorter arrows represent the 1 sigma confidence
errors on this direction.
\label{fig3}}
\end{figure}

\begin{figure}
\centering
\includegraphics[angle=0,scale=.80]{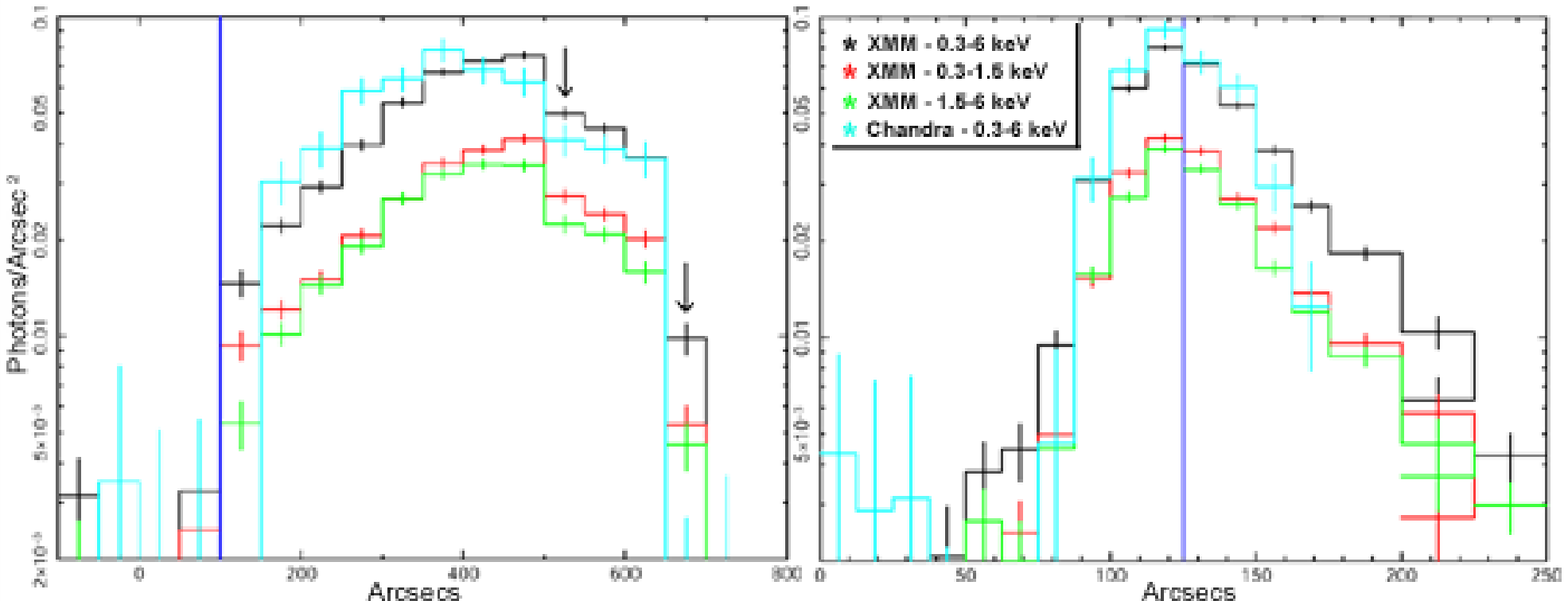}
\caption{Background-subtracted surface brightness profiles of the tail along its main axis (APM, left panel) and orthogonal to it
(OPM, right panel). The PN plus MOS profiles in the 0.3-6, 0.3-1.5 and 1.5-6 keV energy ranges are marked, respectively, with black,
red, and green. The cyan line shows the smoothed 0.3-6 keV {\it Chandra} brightness profile. This last has been convolved with a 
Lorentzian kernel of $\Gamma$ = 7.125", corresponding to a
Full width at half maximum (FWHM) of the distribution of 6.6", the nominal FWHM of the {\it XMM-Newton} mirrors. The integral of the {\it Chandra} profile
has also been normalized to the integral of the 0.3-6 keV {\it XMM-Newton} one in order to show any possible discrepancy. The blue vertical
lines mark the position of the pulsar. Sources 4 and 5, indicated with an arrow, have been removed, possibly enhancing
the decrease in brightness. The flat maximum along the main axis is clearly visible between $\sim$250" and 500" as well as the asymmetrical
shape of the tail in the direction orthogonal to it.
The profiles in the soft and hard X-ray bands are almost the same, with a possible exception right behind the pulsar, where the soft band dominates.
\label{fig4}}
\end{figure}

First, we produced brightness profiles of the nebula along the pulsar motion (APM) and orthogonal to it (OPM).
In order to maximize the statistical significance of each bin, we
carefully evaluated the bin size and the width of extraction
boxes, choosing a scale of 90"$\times$50" APM and 450"$\times$12.5" OPM (see Figure~\ref{fig3}).
For each instrument, we excluded the emission region surrounding the
pulsar, as well as all the serendipitous sources and CCD gaps,
according to our source PSF and calibration studies. Area calculations for each bin take into account the excluded regions.
The presence of straylight annuli, PN CCD gaps, and numerous
serendipitous sources in the faintest part of the nebula, toward the 
south-west, forced us to choose ad hoc boxes in that region, in order to exclude all the contaminants.\\
The brightness profile for both {\it XMM-Newton} and {\it Chandra} are shown in figure~\ref{fig4}.
In order to take into account the different angular resolution of the
{\it XMM-Newton} instruments, the profiles were convolved with a
Lorentzian kernel of $\Gamma$ = 7.125", corresponding to a 
full-width half maximum (FWHM) of the distribution of 6.6", the nominal FWHM of the
{\it XMM-Newton} mirrors. Figure~\ref{fig4} shows the smoothed {\it Chandra} and {\it XMM-Newton} brightness profiles.
While minor differences are found in the count rate profiles shown in the figure, taking into account both the different (energy-dependent) effective areas, the flux profiles are compatible.
The {\it XMM-Newton} data show a hint of extended emission, at brightness below the {\it Chandra} detection limit,
near the pulsar itself. Such a difference is possibly due to the higher {\it XMM-Newton} sensitivity, that
highlights the fainter part of the nebula, right behind the pulsar.
No variations in the tail profile are detected in the available observations.\\
Along the pulsar motion, the tail emerges from the background within 50" of the pulsar and its flux increases gradually, reaching
a flat maximum between 4' and 7'. Its decrease is faster, though the profile is affected by the two serendipitous AGN (see section ~\ref{ima}) and it
fades below the background level at about 9' from Morla. In the
direction orthogonal to the main axis (OPM), the tail profile is clearly
asymmetric: it shows a fast increase in the
northeast direction (rising to the maximum within 30"), reaches a sharp maximum, in broad agreement 
with the neutron star proper motion direction, then the decrease is slower,
fading at about 1-1.5' in the nearest part of the nebula and 2' in the furthest.
Such a behavior recalls the projection of a cone, with the apex
corresponding to the pulsar position (or somewhere in the vicinity of
the pulsar).\\
Brightness profiles of the tail in the soft (0.3-1.5 keV) and hard
(1.5-6 keV) energy bands show no significant discrepancies, with the
possible exception of the region immediately next to the pulsar
($<$50"), where a hint of a softening in the spectrum is seen.

\subsection{Spectral Analysis} \label{nspe}
\begin{figure}
\centering
\includegraphics[angle=0,scale=.80]{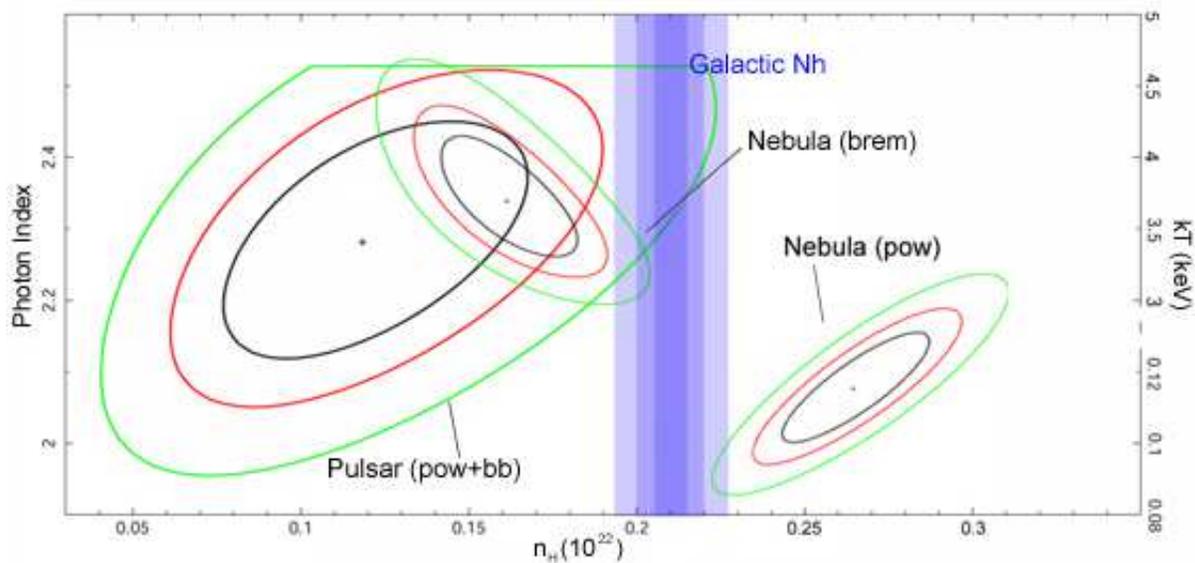}
\caption{Confidence contours from spectroscopy on the pulsar and the tail.
Black, red, and green lines respectively show the error ellipses at
68\%, 90\% and 99\% confidence levels for the absorbing column and the photon index.
The Galactic N$_H$ from the study of serendipitous sources (see section 2.1) is
also shown with errors at 1, 2, and 3 $\sigma$ indicated by different shades of blue.
It is apparent that the Galactic value of N$_H$ and the value obtained by fitting
the powerlaw spectrum on the tail are only marginally compatible.\label{fig5}}
\end{figure}

While in the previous {\it Chandra} observations the spectral analysis of the
tail emission was hampered by the low signal-to-noise ratio (S/N) - $\sim$57\%
of the total counts coming from the background, - the exceptional sensitivity
of {\it XMM-Newton} allows for a deep study of the nebular emission, with a
background contribution of $\sim$35\% .\\
In order to maximize the S/N, we extracted the total spectrum from the brightest
part of the nebula, an elliptical region centered at R.A., Dec. (J2000) 03:58:06.59,
+32:02:05.42, a semi-major axis of 4' and a semi-minor axis of 50".
We excluded from the analysis region a circle with 20" radius around source 5
and the gaps in the CCDs. We extracted the background from source-free circular
regions uncontaminated by the extended emission and located on the same CCD as
the tail. We generated the response and effective area files using the SAS
tools {\tt rmfgen} and {\tt arfgen}. {\it Chandra} data were treated in the
same way as in \citet{del11} and the resulting spectrum was fitted together with the {\it XMM-Newton} ones.
The spectrum of the nebula is well described ($\chi^2_{\nu}$ = 1.06, 162 dof)
by a non-thermal emission model (powerlaw with a photon index
$\Gamma$ = 2.07$\pm$0.08), absorbed by a column
N$_H$ = (2.61$\pm$0.23) $\times$ 10$^{21}$ cm$^{-2}$. Figure~\ref{fig5} shows
confidence contours for N$_H$ and $\Gamma$ for the diffuse feature and for the
pulsar. The value of N$_H$ obtained for the tail seems to be in conflict (at the $\sim3\sigma$
level) with both the one obtained for the pulsar, as well as with the Galactic value,
obtained by fitting the spectra of the serendipitous AGNs in
the field (see section ~\ref{ima}). Such a discrepancy between the values of
N$_H$ led us to consider alternative models. The spectrum of the tail is well
described ($\chi^2_{\nu}$ = 1.06, 162 dof) by a thermal bremsstrahlung model
with a temperature kT = 3.75$_{-0.40}^{+0.48}$ keV, absorbed by the more
realistic column N$_H$ = (1.72$\pm$0.16) $\times$ 10$^{21}$ cm$^{-2}$.
Other similar thermal models such as {\tt mekal} (hot diffuse gas 
\citet{mew85}), {\tt nei} (collisional plasma, non-equilibrium, constant
temperature \citet{bor01}), and {\tt pshock} (plane-parallel shocked plasma,
constant temperature \citet{bor01}) with redshift 0 fit equally well the
spectrum of the tail and require a very low metal abundance.
Table~\ref{tab-2} reports the results of the fit for all the models.

In order to study the spatial-spectral variability of the nebula, we divided
the elliptical extraction region using the two axes, then we fitted both the
absorbed powerlaw and bremsstrahlung models (see Table ~\ref{tab-2}) for each
semi-elliptical region. In this way we obtained spectra from the
semi-elliptical regions nearest to and furthest from the pulsar, and those at
the south-west and north-east. No significant ($3\sigma$) variation was
detected down to $\sim$0.2 in $\Gamma$ or $\sim$1 keV in kT.

\begin{table}
\begin{center}
\caption{Parameters of different X-ray spectral models of the tail\label{tab-2}}
\begin{tabular}{crrrrrr}
\tableline\tableline
Model & $\chi^2_{\nu}$ & N$_H$ & $\Gamma$ & kT & Abund$^a$ & Flux$^b$\\
 & & 10$^{21}$ cm$^{-2}$ & & keV & & 10$^{-13}$ erg cm$^{-2}$ s$^{-1}$\\
\tableline
tot-pow & 1.06 & 2.61$\pm$0.23 & 2.07$\pm$0.08 & - & - & 6.65$\pm$0.40\\
tot-brem & 1.06 & 1.72$\pm$0.16 & - & 3.75$_{-0.40}^{+0.48}$ & - & 4.95$\pm$0.30\\
tot-mek & 1.13 & 1.73$\pm$0.16 & - & 3.76$_{-0.38}^{+0.49}$ & $<$0.099 & 4.57$\pm$0.28\\
near-brem & 1.15 & 1.76$\pm$0.14 & - & 4.72$_{-0.69}^{+0.89}$ & - & 2.23$_{-0.34}^{+0.20}$\\
far-brem & & & - & 3.39$_{-0.33}^{+0.38}$ & - & 2.05$\pm$0.16\\
sw-brem & 1.33 & 1.72$\pm$0.14 & - & 4.01$_{-0.52}^{+0.64}$ & - & 1.83$\pm$0.16\\
ne-brem & & & - & 4.23$_{-0.46}^{+0.55}$ & - & 2.48$\pm$0.18\\
\tableline
\end{tabular}
\tablenotetext{a}{Metal abundances (He fixed at cosmic). The elements included are C, N, O, Ne, Mg, Si, S, Ar, Ca, Fe, Ni, see \citet{and89}}
\tablenotetext{b}{Unabsorbed 0.3-10 keV flux.}
\tablecomments{Here we report the best fits of models that can describe the nebular emission. The elliptical 
extraction region was chosen after an accurate study based on the signal-to-noise ratio. We used
an absorbed powerlaw, bremsstrahlung and {\tt mekal}, respectively. Thermal models obtain similar
temperatures, absorption columns and abundances. The last four spectra describe the emission
from the nearest and the furthest semi-elliptical nebular regions from the pulsar,
the south-west and north-east ones, respectively, using a bremsstrahlung model to fit data. To describe the total
tail emission, we simultaneously fitted the {\it XMM-Newton} spectra and the {\it Chandra} ones. For the
different parts of the tail we used only {\it XMM-Newton} spectra (due to the different PSFs
of the two telescopes) and we simultaneously fitted the furthest with the nearest nebular spectra
and the south-east with the north-west ones, chaining only the N$_H$ values.
All the errors are at a 90\% confidence level.}
\end{center}
\end{table}

\section{Discussion} \label{disc}

\subsection{The Pulsar}
The X-ray spectrum of the Morla pulsar is a combination of thermal radiation from hot
spot(s) superimposed on a non-thermal power-law component. A straight estimate
of the NS polar cap size, based on a simple centered dipole magnetic field
geometry (see e.g. \citet{del05}) and a standard NS radius of 10 km, predicts
in this case a polar cap radius of about 85 m. The observed emitting radius
is about 450 m, significantly smaller than the entire surface of any reasonable
NS but definitely bigger than the expected polar cap. This is true both using the blackbody model and the NS atmosphere model,
although the latter yields a larger emitting area and a lower temperature.
Such a large size of the spots has
also been found in other well-known pulsars, like PSR B0656+14 and PSR B1055-52
\citep{del05}, and it is still poorly understood. The luminosity associated with
the Morla polar cap is 1.1 $\times$ 10$^{30}$ erg s$^{-1}$, a factor of a few lower
than the luminosity expected from polar caps heated by e$^{+/-}$ currents.
However, it is consistent with the expectations from polar cap heating due to
the bombardment by particles created only by inverse Compton scattered photons
\citep{har01,har02}. PSR J0357+3205 is close to the death line for production
of e$^{+/-}$ pairs by curvature radiation photons and this could explain the
reduced polar cap heating.\\
While the majority of X-ray emitting isolated neutron stars (e.g. \citet{kas06}),
exhibit thermal emission coming from the whole NS surface, no such a component
is required to fit the {\it XMM-Newton} spectrum.
Using the blackbody model, for a NS radius of 10 km and a distance of 500 pc,
the 3$\sigma$ upper limit on the temperature is 4.4 $\times$ 10$^5$ K. This
constraint makes Morla the coldest NS in its age range, slightly cooler than
the coeval radio-quiet pulsar RX J1856-3754 (e.g. \citet{sar12}). Such a
difference can be attributed to different magnetic fields, and thus different
magneto-thermal evolution (e.g. \citet{pon09}). The non-thermal X-ray
luminosity of Morla is in broad agreement with the relation between the X-ray
luminosity of rotation-powered pulsars and their spin-down luminosity
\citep{pos02,kar08,mar11,mar12}.

\subsection{The Tail} \label{tail}

The morphology of the tail, its spatial extension - $\sim$1.3 pc at 500 pc
assuming no inclination with respect to the plane of the sky ($i$), - and the lack of any
other related Galactic source led De Luca et al. 2011 to associate
unambiguously the nebular emission with the pulsar. Elongated X-ray tails
coupled with pulsars are quite common (e.g. \citet{kar08a}) and are
usually interpreted within the framework of bow-shock, ram-pressure-dominated
pulsar wind nebulae (e.g. \citet{gae06}). When a pulsar moves supersonically,
the shocked pulsar wind is expected to flow in a region downstream
of the termination shock (the cavity in the interstellar medium created by the
moving Neutron Star and its wind), confined by ram pressure. X-rays are produced by
synchrotron emission from the wind particles accelerated at the termination
shock, which is typically seen (if it can be resolved) as the brightest part
of the extended structure (see, e.g., \citet{kar08b}), according to the
expectations from magnetohydrodynamic simulations \citep{buc02,van03,buc05}.
\citet{del12} have detected and measured a very large proper motion for the
pulsar: 165$\pm$30 mas yr$^{-1}$ - 390 km s$^{-1}$ across the plane of the sky,
at 500 pc - in the direction opposite to the tail, thus compatible with such
a scenario.

The tail shows no other similarities with other, synchrotron-emitting nebulae
(for a complete discussion about Morla's synchrotron-emitting model see
\citet{del11}).
The first problem arises from energetic requirements for the emitting particles.
According to \citet{gol69}, the maximum potential drop between the pole and the
light cylinder (in an aligned pulsar) is
$\Delta\Phi$ = (3 $\dot{E_{rot}}$/2c)$^{1/2}$. In the Morla case, the electrons are
accelerated up to a maximum Lorentz factor $\gamma_{max}\sim$10$^8$, which can
be considered as an upper limit for the electrons injected in the pulsar wind
nebula. The low ${\dot E_{rot}}$ of Morla requires an ambient magnetic field
as high as $\sim$50 $\mu$G - while the mean Galactic value is $\sim$1 $\mu$G
(\citet{jan12} and references therein). However, one must also take into account
the magnetic field carried by the particles themselves. An independent estimate
of the magnetic field in the tail can be obtained from the measured synchrotron
surface brightness \citep{pav03}. Taking into account the uncertainty on the
parameters we can conclude that a reasonable value of the magnetic field in the
tail is in the range $20-100$ $\mu$G,
consistent with the value of $\sim$50 $\mu$G obtained before.

The second problem is the lack of diffuse emission surrounding the pulsar,
where the emission from the wind termination shock should be brightest, as
observed in all the other known cases (e.g. \citet{gae04,mcg06,kar08}).
Such a shock could be unresolved by {\it Chandra} only in the case of an
extremely large ambient density (several hundred atoms per cm$^3$)
and/or a pulsar speed higher than some thousands km s$^{-1}$. Taking into
account the measure of the proper motion, the distance of the pulsar should
be larger than 1.3 kpc or the inclination of the motion should be higher
than 70$^{\circ}$. Such conditions imply a tail at least 3.5 pc long,
so that it becomes even more difficult to account for the energy of the particles.
This picture implies also that a significant fraction of the point-like flux
comes from a non-thermal termination shock, reducing dramatically the power-law
component of the pulsar. While it is acceptable
for the non-thermal component of the pulsar to be a factor of some smaller than the value we obtained
\citep{mar11,mar12}, the lack of nebular emission around the pulsar, where
the synchrotron emission should be stronger, remains unexplained.

The brightness profile along the main axis of the tail is consistent with no emission
within 30-50" of the pulsar, then it slowly increases to reach a broad
peak at $\sim$4'. This behavior is remarkably different from that observed
in all the known synchrotron nebulae that show peaks close to the position of
their parent pulsar (e.g. \citet{kar08}), where the acceleration of the wind
particles is the most important. It is very hard to explain in this scenario why
the emission peaks at $\sim$1 pc from the pulsar.\\
The asymmetric brightness profile of the tail in the direction perpendicular
to its main axis is characterized by a sharp northeastern edge and a slow
decay toward the south-west direction. The synchrotron emission from the nebula is
only marginally dependent on the ISM density, so that 
ad-hoc and large variations in the ISM are demanded, in order to explain
this strange profile. Even if possible, such a case is hardly believable.\\
Furthermore, the inferred synchrotron cooling time implies a significant
spatial-spectral evolution with respect to the distance from the pulsar,
as observed in all synchrotron nebulae \citep{kar08}. Thanks to
the unique throughput of {\it XMM-Newton}, we could perform a deep
spatial-spectral study of the tail, analyzing separately the furthest and
the nearest parts of the nebula. No significant differences in the spectral
index are detected (see section ~\ref{nspe}). The lack of
variation cannot be easily explained in terms of a synchrotron nebula. The lack of such
a softening would require some particle re-acceleration mechanism within the tail
itself.\\
The excellent throughput of the {\it XMM-Newton} instruments has
revealed another problem of the synchrotron nebula scenario, connected to
the value of the column density (N$_H$ ) obtained by fitting the non-thermal
model of the nebula. While it is marginally compatible with the value
obtained for the pulsar, the inferred N$_H$ appears to be
larger than the Galactic one in that
direction, measured by fitting serendipitous AGNs in the field (see
Figure~\ref{fig5}).

As reported in section~\ref{nspe}, a thermal bremsstrahlung model fits
equally well the spectrum of the tail and produces an estimate of N$_H$
in agreement with that of the pulsar and the Galactic one. In the bremsstrahlung
scenario, the tail emission arises from the shocked ISM material heated
up to X-ray temperatures. Under the assumption of a strong shock-wave and
from the junction conditions at the shock front, we can estimate the
speed of the pulsar directly from the temperature of the shocked
ISM material at the head of the bow shock:
$v_{psr}=\sqrt{16kT/(3\mu m_p)}$, where $\mu$ is the molecular weight of
the ISM gas and we assumed an adiabatic index $\gamma_{ad}=5/3$.
The temperature obtained by fitting the tail spectrum with a thermal
bremsstrahlung model, $kT\approx3.75$ keV, implies a pulsar speed
$v_{psr}\simeq1376\mu^{-1/2}$ km s$^{-1}$. From the values of the abundances
resulting from the spectral fit, we can assume that $\mu\simeq0.54$, so
the speed of the pulsar is about $1900$ km s$^{-1}$, which would make it the largest ever
observed in a pulsar. If we combine this result with the pulsar proper
motion we can estimate the inclination angle $i$ of the pulsar motion
(and thus of its tail), $i\approx$ acos($0.2 \times d_{500}$).
Moreover, through the value of the photon flux measured using a thermal
bremsstrahlung model, it is possible to estimate the number density of
the shocked ISM gas as $n\simeq0.86d_{500}^{-1/2}$ atoms cm$^{-3}$.

Thermal bremsstrahlung emission acts as a cooling mechanism. The time
scale for cooling can be estimated as $t_{cool}\simeq(E_{thermal}/J)$,
where $E_{thermal}\simeq nkT$ is the thermal energy per unit volume and
$J$ is the volume emissivity. This approximation gives
$t_{cool}\simeq6\times10^3T^{1/2}(n\bar{g}_{B})^{-1}$ yr, where
$\bar{g}_{B}$ is the frequency-averaged Gaunt factor that typically
ranges between $1.1$ and $1.5$, depending on the plasma temperature
\citep{pad00}. From our estimates of the temperature and number density
of the shocked ISM gas we expect $t_{cool}\simeq10^7$ yr. Considering
a pulsar proper motion of 165$\pm$30 mas yr$^{-1}$ and an extension of
the nebula of about $9$', then the age of the tail ($\sim$3300 yr)
is much lower than its cooling time. This result can explain why we have
not measured a significant evolution of the temperature of the tail as a
function of the distance from the pulsar. On the other hand, the shorter than expected
length of the tail can be explained assuming an ad-hoc decrease of a $\sim$3 factor in the 
number density $n$ beyond the end of the tail - the volume emissivity
of the thermal bremsstrahlung emission is $J\varpropto n^2T^{1/2}$.
Such a strong dependence on the ISM density can explain the asymmetrical
shape of the nebula in the direction orthogonal to its main axis: this
could be a consequence of an inhomogeneity of the number density in the
nebula, quite typical in the ISM.

From the conservation of mass, energy, and momentum on the two sides
of the shock front and from the values measured in the tail (e.g. the pre-shock
temperature $T_0$ and number density $n_0$), we can calculate
the same physical quantities outside the nebula:
\begin{equation}
 \frac{n}{n_0}=\frac{(\gamma+1)M^2}{(\gamma+1)+(\gamma-1)(M^2-1)}
\end{equation}
\begin{equation}
 \frac{T}{T_0}=\frac{\left[(\gamma+1)+2\gamma(M^2-1)\right]\left[(\gamma+1)+(\gamma-1)(M^2-1)\right]}{(\gamma+1)^2M^2}
\end{equation}
where $M=1/\sin(\alpha)$ is the Mach number and depends on the cone
angle of the tail ($\alpha$) and an adiabatic index $\gamma_{ad}=5/3$
is assumed for the ISM gas. In the strong shock-wave limit ($M>>1$)
$n_0=n/4\simeq0.22d_{500}^{-1/2}$ atoms cm$^{-3}$ while $T_0$ depends
on the Mach number and thus on the inclination angle $i$.
From the geometry of the cone angle of the nebula, a plausible value
of $T_0$ is in the range of $10^5$-$10^6$ K, for $i \gtrsim 60^{\circ}$.
The pre-shock temperature is consistent with that of the hot phase
of the ISM, which fills a large fraction of the Galaxy (e.g.
\citet{bla00}). Such a scenario is also in full agreement with the
requirement of a full ionized ISM by \citet{del12}.
We note that the density required by our model is
slightly higher than expected for the hot ISM: this may result
from the concerted action of massive stellar
winds and supernova explosions.
\citet{yao05} report the presence of dense
(10$^{-3}$-1 atoms cm$^{-3}$), hot ($\sim$10$^6$ K) envelopes
of gas in our Galaxy. On the other hand, thermal models such as
{\tt mekal}, {\tt nei}, and {\tt pshock} point to a very low
metallicity of the ISM around the pulsar, making more difficult
to explain this dense envelope as the result of a supernova
explosion or stellar winds. Future multi-wavelength observations
of the field and more detailed Galactic gas models could better
constrain and explain the ISM composition around Morla.

A shock-wave scenario can easily explain the lack of diffuse
emission surrounding the pulsar. Most of the energy in the
pre-shock flow is carried by the ions - the kinetic energy in
the streaming of the electrons is less than $1/2000$ of the
total - but the electrons are generally responsible for the
cooling of the plasma: this means that the electron temperature
($T_e$) is responsible for the dynamics of the shock and its 
emitted spectrum. The electrons must be heated by the ions
before the emission becomes detectable and this process takes
some time. Coulomb heating of the electrons behind the shock
proceeds at a rate
\begin{equation}
 \frac{dT_e}{dt}=\frac{T_i-T_e}{t_{eq}}
\end{equation}
where $T_i$ is the ion temperature and
$t_{eq}=7.7(T_e^{3/2}/n)$ s $\sim$70 kyr is the equipartition
time for a fully ionized plasma with a Coulomb logarithm of 30
(where $T_e$ is in units of K and $n$ in cm$^{-3}$). If all the
post-shock energy is in the ions ($T_e\ll T_i$), then the
increase in T$_e$ due to collisional heating follows:
$T_e(coll)=1.7\times10^7(n_0t_3v_{psr8}^2)^{2/5}$ K, where
$t_3$ is the age of the tail in units of $10^3$ yr,
$n_0$ is the pre-shock number density and $v_{psr8}$ is the
pulsar speed in units of $10^8$ cm s$^{-1}$. Thus
 $T_e(coll) \lesssim 6.6\times10^7d_{500}^{-1/5}$ K, not
significantly different from the value observed in the tail.
Moreover, $T_e(coll)\lesssim5.5\times10^{6}$ K within $\sim30$"
of the pulsar (corresponding to $\sim$200-300 years, from
the measure of the proper motion). This fully explains the
lack of diffuse emission surrounding the pulsar.

So far, no evidence of other pulsars characterized
by a thermal bremsstrahlung emitting tail has been reported. This could be explained
by the peculiar conditions it requires, e.g. very fast pulsars and an hot
and dense ISM. With its estimated velocity of 1900 km s$^{-1}$,
J0357+3205 is the fastest moving pulsar known. Among all the pulsars listed in
the ATNF Pulsar Catalogue \citep{man05}, only 2 have a very high velocity
($\gtrsim 700$ km/s), reliably measured through an estimate of the distance
obtained by other methods than the dispersion measure.
PSR B2224+65, powering the Guitar Nebula \citep{cor93} and showing an X-ray
tail similar to Morla \citep{hui07}, has a speed of at least $865$ km s$^{-1}$
\citep{man05}. J0821-4300, the central compact object in the Puppis A supernova
remnant has a speed possibly higher than 1000 km s$^{-1}$ \citep{hui06,win07,bec12}.
The spin-down power of PSR B2224+65 is similar to that of Morla
($\dot{E}_{rot}$=$5.9\times10^{33}$ erg s$^{-1}$ and
$1.2\times10^{33}$ erg s$^{-1}$), but the tails are different.
While both tails have a similar brightness profile and show no spectral
variations, the tail of PSR B2224+65 is misaligned (118$^{\circ}$) with respect to
the proper motion of the pulsar. Alternative explanations have been proposed:
according to \citet{ban08} the feature could be produced by highly energetic
particles which escaped the pulsar wind termination shock and are confined
by an interstellar magnetic field. This scenario accounts for a faint
counter-feature, detected on the opposite side of the pulsar with respect to
the tail, but does not explain the absence of spectral variations along the
tail \citep{joh10}. Another possible scenario is that the observed emission
comes from a ballistic jet, similar to those of AGNs \citep{joh10}, but also
this picture has some problems.\\
The key to understand why these two pulsars do not emit through thermal
bremsstrahlung lies in the ISM characteristics. This emission is detectable
in X-rays only if both the ISM density and temperature are high enough
($\sim$0.2 atoms cm$^{-3}$ and $\sim$10$^5$-10$^6$ K). It is hardly
believable that the ISM around a CCO like J0821-4300, lying inside its
supernova remnant, could be denser than $\sim$10$^{-3}$ atoms cm$^{-3}$:
this can explain the lack of a bremsstrahlung tail. In the case of
PSR B2224+3205, the ISM density, estimated from the H$\alpha$ bow-shock
emission to be $<1$ cm$^{-3}$ \citep{cha02}, is lower than the ISM density
predicted around Morla. No constraints on the ISM temperature around
PSR B2224+3205 appear in the literature: it is not excluded that a low
temperature of the ISM may contribute to the lack of a bremsstrahlung tail.

Recently, \citet{tom12} studied the case of the X-ray source IGR J11014-6103,
suggesting an association with the supernova remnant MSH 11-61A
\citep{kes68}, thus inferring a velocity of 2400-2900 km s$^{-1}$.
IGR J11014-6103 shows a complex morphology, consisting of a point source,
an elongated tail $\sim50$" long, aligned to the direction of the proper
motion, and a feature of $\sim$4', nearly perpendicular to the proper motion.
Spectral and spatial analyses suggest that IGR J11014-6103 could be a pulsar,
therefore providing a natural explanation for the co-aligned feature as
a bow-shock ram-pressure-confined pulsar wind nebula. Moreover, the feature
resembles the tail of Morla: its luminosity peaks far from the pulsar and
only a few photons are detected near the pulsar. The low statistics prevent
us from studying further the similarities between the two tails. However, if
the association with MSH 11-61A is confirmed, its distance would be 10 kpc,
implying a tail of $\sim2$ pc, comparable to that of Morla.

\subsection{The Distance and inclination}

Our deep {\it XMM-Newton} observation of the field of J0357+3205 allowed us
to better constrain the distance of the pulsar using the interstellar
absorption value. The Galactic absorption column in the direction of the 
pulsar, predicted by \citet{dic90} at $\sim$500 pc based on the HI
distribution, is fully compatible with our best fit value for the pulsar
and the tail, pointing to a lower limit for the pulsar distance of
$\sim$300 pc. This result is in broad agreement with the value 
estimated by scaling the $\gamma$-ray flux of the pulsar
\citep{saz10}: the method hinges upon the observed correlation
between the intrinsic $\gamma$-ray luminosity and spin-down power of the
pulsar \citep{abd10a}, assuming a beam correction factor of 1 for the
$\gamma$-ray emission cone of all the pulsars \citep{wat09}.
By applying this relation, we have a $\gamma$-ray efficiency of 0.33
and a distance of $\sim$500 pc. We obtain a new upper limit of $\sim$900 pc
by requiring the $\gamma$-ray efficiency to be less than 1.\\
Another important piece of information could come from the serendipitous
stars we found in the field: their distance estimates from optical analyses
could provide lower limits on the distance of the pulsar. Future
multi-filter optical observations will provide this limit.\\
Under the hypothesis of a bremsstrahlung-emitting tail, we infer a pulsar
velocity of $\sim$1900 km s$^{-1}$. The estimated value of the pulsar
proper motion, 165$\pm$60 mas yr$^{-1}$, allows us to set another upper
limit for the pulsar distance, being 2.4$_{-0.6}^{+1.4}$kpc for a purely
transversal motion. However, as reported in section ~\ref{tail}, a null
inclination leads to unrealistic values ($\sim$10$^7$ K) of the ISM
temperature, so that an inclination angle larger than 60$^{\circ}$
is required. This lowers the upper limit to 1.2$_{-0.3}^{+0.7}$ kpc,
still less constraining than that obtained by computing the pulsar
efficiency.

For a pulsar with a kick velocity of $\sim$1900 km s$^{-1}$, a
proper motion of 165 mas yr$^{-1}$ and a distance of
300 pc $<D<$ 900 pc, the inclination of the system with respect to
the plane of the sky ranges between 68$^{\circ}$ and 83$^{\circ}$, the density
of the ISM between 0.15 and 0.3 atoms cm$^{-3}$, and its temperature
between 1 and 9 $\times$ 10$^5$ K. The length of the tail is
$\sim$6.5 pc, independently of the distance.

Although the study of the kinematics of the pulsar is beyond the
scope of this paper, the small component of the proper motion
perpendicular to the Galactic plane points to a birth site outside
the Galactic plane, suggesting that its progenitor might have been
a runaway star, as already introduced in \citet{del11}.


\section{Conclusions} \label{conc}

We have characterized the properties of the middle-aged
$\gamma$-ray and X-ray pulsar J0357+3205 (Morla) and its tail.
The X-ray spectrum of the pulsar is consistent with a magnetospheric
non-thermal component plus thermal emission from hot spots;
the lack of any detected thermal component from the whole surface
makes Morla the coldest NS in its age range.\\
Folding  the X-ray photons at the expected pulsar period using the LAT ephemeris yields
marginal evidence for modulation, with 2 peaks,
neither of them aligned with the single gamma-ray peak (see
Figure~\ref{figp}).

We confirmed the presence of a diffuse emission $\sim$9' long,
the largest tail of X-ray emission associated with any rotation-powered
pulsar. Such an extended emission cannot be explained in terms
of the usual bow-shock ram-pressure-dominated pulsar wind nebula.
In fact, the following problems arise:\\
- the existence of the tail is problematic for a pulsar with such a low
$\dot{E}$ as Morla;\\
- no bow shock has been resolved around the pulsar;\\
- the lack of any nebular emission around the pulsar, where the particle
acceleration is maximum, cannot be explained;\\
- the nebular spectrum lacks any spatial evolution, a necessary signature
of the radiative cooling of the electrons accelerated at the wind
termination shock;\\
- the very asymmetric brightness profile in the direction perpendicular
to the main axis of the tail requires a large ad-hoc inhomogeneity of
the ISM around the pulsar;\\
- fitting the {\it XMM-Newton} nebular spectra with a powerlaw, we found
the nebular N$_H$ to be higher than the Galactic value and only marginally
in agreement with the N$_H$ of the pulsar.

We propose a thermal bremsstrahlung model as an alternative explanation
of the tail emission. In this scenario, the emission comes from the
shocked ISM material heated up to X-ray temperatures. This model gives
full account of the peculiar features of the tail:\\
- the lack of any detectable spatial evolution in the tail spectrum is due
to the long bremsstrahlung cool-down time;\\
- the peculiar asymmetries of the brightness profile can be interpreted in
terms of small changes in the ISM density that strongly affect the
brehmsstralung emissivity;\\
- most of the energy in the pre-shock flow is carried by the ions, while
the electron temperature is responsible for the X-ray emission; the Coulomb
heating time of the electrons behind the shock is fully in agreement with
the lack of any detected nebular emission near the pulsar;\\
- the value of N$_H$ measured in the tail agrees both with the value
obtained for the pulsar and the Galactic value.\\
This scenario allows us to estimate some parameters of the pulsar and
of the ISM around it. For a bremsstrahlung-emitting tail we estimate
the pulsar distance to be between 300 and 900 pc. A pulsar velocity of
$\sim$1900 km s$^{-1}$ is required, in agreement with the pulsar
proper motion for distances of some hundreds parsecs and a high inclination.
The mean density of the ISM is required to be $\sim$0.2 atoms cm$^{-3}$
and the temperature of some 10$^5$ K. This type of hot gas usually
presents lower densities, but a denser phase (possibly detected by
\citet{yao05}) is predicted to be the result of the action of massive
stellar winds and supernova explosions. However, the low metallicity
we obtained from the spectral fit makes the explanation of this gas
envelope more diffcult.\\
A high inclination of the system ($>$70$^{\circ}$) is predicted in both
models. This value provides further support to the bremsstrahlung model,
where the cool-down time demands for a longer tail, and adds to the
problems of a synchrotron model.

For all these reasons, we believe Morla's nebula to be the first example
of a new ``turtle's tail'' class of thermally-emitting nebulae, ironically
associated with fast moving pulsars.\\
Until now we have no clear evidence of other pulsars characterized
by a tail emitting via thermal bremsstrahlung, possibly for the
requirements of a very fast pulsar and a hot, dense ISM. Moreover,
energetic pulsars can also generate classic synchrotron nebulae, that
may outshine a bremsstrahlung component. The recently discovered
nebula associated with the source IGR J11014-6103, suspected of being a
high-velocity pulsar, resembles the tail of Morla: further studies
of this feature could lead to a second example of a turtle's tail nebula.




\acknowledgments

This work was supported by the contract Swift ASI-INAF 1-004-11-0. We warmly thank the ApJ referee
for the useful comments and suggestions as well as the IASF-INAF staff who constructively
participated in the discussion during my ``Astrosiesta'' presentation.



{\it Facilities:} \facility{CXO (ACIS)}, \facility{XMM (EPIC)}, \facility{Mayall (MOSAIC,FLAMINGOS)}





\clearpage


\end{document}